\documentclass[conference]{IEEEtran}
\IEEEoverridecommandlockouts
\usepackage{cite}
\usepackage{amsmath,amssymb,amsfonts}
\usepackage{algorithmic}
\usepackage{graphicx}
\usepackage{tabularx}
\usepackage{romannum}
\usepackage{textcomp}
\usepackage{makecell}
\usepackage{multirow}
\usepackage{xcolor}
\def\BibTeX{{\rm B\kern-.05em{\sc i\kern-.025em b}\kern-.08em
    T\kern-.1667em\lower.7ex\hbox{E}\kern-.125emX}}
\begin{document}

\title{Synthetic Waveform Generation for Satellite, HAPS, and 5G Base Station Positioning Reference Signal Using QuaDRiGa
\thanks{This work was supported in part by Discovery Grants RGPIN-2017-06261 and RGPIN-2022-05231 from the Natural Sciences and Engineering
Research Council of Canada (NSERC), in part by the Coordenação de Aperfeiçoamento de Pessoal de Nível Superior - Brazil (CAPES) - Finance Code 001, in part by CNPQ, and in part by FAPERJ.}
}

\author{\IEEEauthorblockN{Hongzhao Zheng\textsuperscript{\dag}, Mohamed Atia\textsuperscript{\dag}, Halim Yanikomeroglu\textsuperscript{\dag}, Paulo S. R. Diniz\textsuperscript{\S}}
\IEEEauthorblockA{\textsuperscript{\dag}\textit{Department of Systems and Computer Engineering, Carleton University, Ottawa, Canada}\\
\textsuperscript{\S}\textit{Department of Electronics and Computer Engineering, Universidade Federal do Rio de Janeiro, Rio de Janeiro, Brazil}\\
Emails: \{hongzhaozheng, halim\}@sce.carleton.ca, mohamedatia@cunet.carleton.ca, diniz@smt.ufrj.br}\\
}
\maketitle

\begin{abstract}
Waveform generation is essential for studying signal propagation and channel characteristics, particularly for objects that are conceptualized but still need to be operational. We introduce a comprehensive guide on creating synthetic signals using channel and delay coefficients derived from the Quasi-Deterministic Radio Channel Generator (QuaDRiGa), which is recognized as a 3GPP-3D and 3GPP 38.901 reference implementation. The effectiveness of the proposed synthetic waveform generation method is validated through accurate estimation of code delay and Doppler shift. This validation is achieved using both the parallel code phase search technique and the conventional tracking method applied to satellites. As the method of integrating channel and delay coefficients to create synthetic waveforms is the same for satellite, HAPS, and gNB PRS, validating this method on synthetic satellite signals could potentially be extended to HAPS and gNB PRS as well. This study could significantly contribute to the field of heterogeneous navigation systems.
\end{abstract}


\section{Introduction}
Signal waveform generation is crucial in signal processing and various aspects of wireless communications. This is particularly true for emerging and yet-to-be-deployed communication carriers or signals, including low Earth orbit (LEO) satellites, high-altitude platform stations (HAPS), and 5G new radio (NR) positioning reference signal (PRS). HAPS, which are stratospheric platforms situated approximately 20 km above the Earth's surface, are capable of delivering communication services~\cite{b4}. PRS, on the other hand, is a type of signal specifically designed for accurate location and positioning services and is expected to be integrated into 5G base stations, also known as gNodeBs (gNBs)~\cite{b16}. Despite the wide array of commercial waveform generators available, the rapid pace of research often leaves these tools lacking certain advanced features like object creation with a customizable location and realistic multipath simulation, in addition to the expensive cost for purchasing the hardware. Consequently, the community tends to favor software-defined radio (SDR), where radio elements are realized through software for its cost-effectiveness, adaptability, and up-to-date capabilities~\cite{b5}.

Generally, an SDR system includes a radio frequency (RF) front end coupled with a personal computer (PC) that has a analog-to-digital converter (ADC). The bulk of the signal processing tasks are carried out digitally through software-based radio components on the PC. The goal of SDR technology is to position the ADC as close to the RF front end as possible to enhance the extent of signal processing conducted digitally via software~\cite{b6}. This approach is still considered an SDR system even in the absence of an RF front end, with the signal waveform being generated entirely through software. Furthermore, research in this field often focuses on specific aspects of signal propagation and characteristics. Consequently, researchers may prioritize simpler signals that are quicker to generate when studying certain propagation effects.

Waveform generation can be realized either through simulation and analysis software platforms like Skydel GNSS Simulation Software or coding. Although the performance delivered by software tends to be better than that of coding, the same constraint where software follows closely with standards limits the use cases of software to the trending technologies. Therefore, waveform generation via coding with the assistance of an advanced channel generator, which considers latest communication standards can be an effective solution. The quasi-deterministic radio channel generator (QuaDRiGa), which follows the geometry-based stochastic channel models (GSCM) approach, has been widely acknowledged and adopted for channel generation~\cite{b7,b8,b9,b17,b18,b19,b20}. QuaDRiGa offers several challenging features such as continuous time evolution, spatially correlated large and small-scale fading, transitions between varying propagation scenarios, and so on. In addition the incorporation of a statistical ray-tracing model has made it capable of  modeling channel fading and Doppler effect effectively. 

In this research, we focus on generating synthetic signals that accurately mimic real-world signal transmission behaviors, including aspects like free space path loss, reflection, refraction, scattering, multipath propagation, shadowing, and Doppler shift. Typically, these features are available simultaneously only in computationally intensive ray-tracing models. We detail a method for generating these synthetic signals for satellites, HAPS, and gNBs, utilizing channel and delay coefficients derived from the QuaDRiGa channel generator. This approach builds upon our previous research~\cite{b2}, where we generated channel and delay coefficients for dual mobile space-ground links using QuaDRiGa. In the current study, we have expanded our scope to include satellites, HAPS, and gNBs. We present the Doppler spectrum of these ranging sources to highlight the essential features of the channel information. We conduct basic signal acquisition and tracking tests to assess the accuracy of our generated synthetic signals. Finally, we propose several potential techniques that could enhance signal-tracking performance. The main contributions of this paper are listed as follows:
\begin{enumerate}
    \item We have extended the scope of the system in our QuaDRiGa-based tutorial to include HAPS and gNBs, where the locations for HAPS and gNB are simulated using the Skydel GNSS simulation software. This extension leads to a multi-frequency vertical heterogeneous network (VHetNet) simulation with mobile communication links.
    \item A synthetic waveform generation method based on the channel and delay coefficients from QuaDRiGa is proposed. This can be used to generate any radio frequency (RF) communication waveform.

    \item The case study on the generated synthetic satellite signal validate the proposed waveform generation method. This is evidenced by the accurate estimation of the code delay and Doppler shift using a basic signal acquisition and tracking method.
\end{enumerate}

\begin{figure}[t]
\centerline{\resizebox{\columnwidth}{!}
{\includegraphics{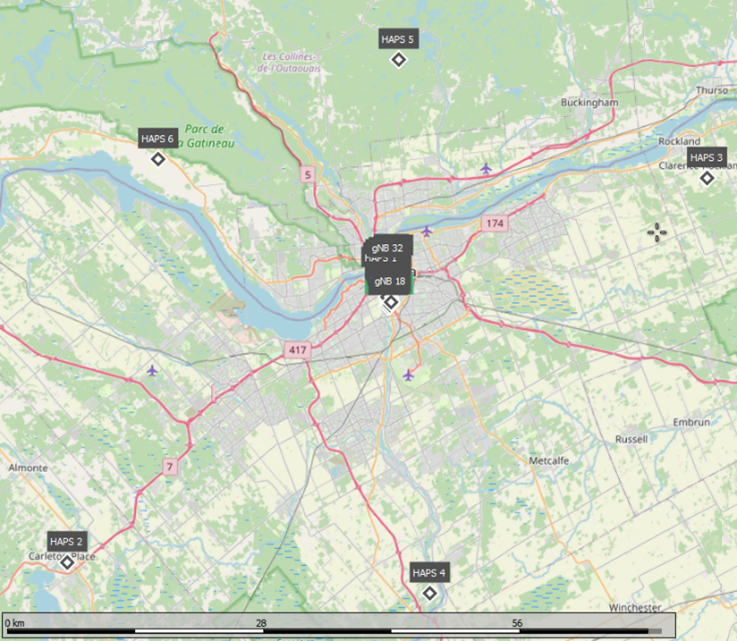}}}
\caption{Locations of simulated HAPS.}
\label{fig5}
\end{figure}

\begin{figure}[t]
\centerline{\resizebox{0.8\columnwidth}{!}
{\includegraphics{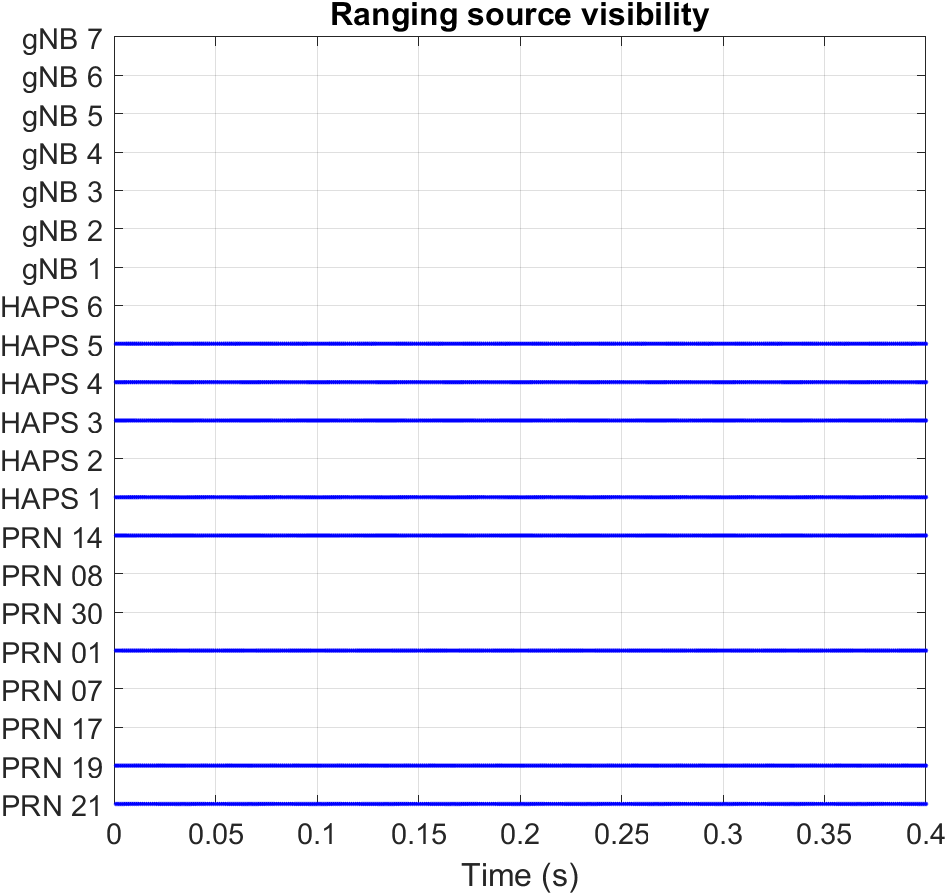}}}
\caption{Visibility plot for satellite, HAPS, and gNB (blue line indicates a LOS scenario).}
\label{fig7}
\end{figure}

\begin{table}[t]
\caption{Simulation Parameters for Channel and Delay Coefficient Generation.}
\begin{center}
\begin{tabularx}{\linewidth}{>{\centering\arraybackslash}X|>{\centering\arraybackslash}X|>{\centering\arraybackslash}X|>{\centering\arraybackslash}X}
\Xhline{1pt}
\textbf{Parameters} & \textbf{Satellite} & \textbf{HAPS} & \textbf{gNB}\\
\Xhline{1pt}
$f_{ch}$ & \multicolumn{3}{c}{\centering 40 kHz}\\
\hline
$T$	& \multicolumn{3}{c}{\centering 0.4 s}\\
\hline
$P_{TX}$ & \multicolumn{3}{c}{\centering 0 dBm}\\
\hline
$G_{TX}$ & \multicolumn{3}{c}{\centering 40 dBi}\\
\hline
$f_c$	 & 1575.42 MHz & 1176 MHz & 4 GHz\\
\hline
Antenna & Parabolic dish	& Parabolic dish & 3GPP-3D\\
\hline
$(M,N)$ & N/A	& N/A & (8,1)\\
\hline
Polarization & RHCP	& RHCP & +/-45 deg\\
\hline
$\phi$ & N/A	& N/A & 8 deg\\
\Xhline{1pt}
\end{tabularx}
\label{tab1}
\end{center}
\end{table}

\section{System Model}
In this study, we continue with the system model established in our previous research. We generate the positions of HAPS and gNBs using the Skydel GNSS simulation software~\cite{b2}, while real satellite orbits are incorporated through a two-line element (TLE) file~\cite{b10}. The HAPS in our simulation are placed 20 km above the ground, moving in a circular path with a 300 m radius at a speed of 30 m/s. The height for all gNBs is set to 25 m. With the simulated topology, the minimum distance from any gNB to the vehicle's path is measured to be approximately 50 m. The receiver, traveling at a speed of 50 km/h, is equipped with a patch antenna oriented towards the sky. The positions of the simulated HAPS are depicted in Fig.~\ref{fig5}.

Given the complexity of QuaDRiGa and the diverse infrastructure in this project, we opt for a simulation duration of 0.4 s. To ascertain the presence of a line-of-sight (LOS) connection between a satellite/HAPS and the receiver, we employ the LOS probability model proposed in ~\cite{b13} and ~\cite{b14}. Similarly, for determining LOS between a gNB and the receiver, we use the LOS probability model for the Urban Macro (UMa) scenario as outlined in 3GPP TR 38.901 Release 16~\cite{b15}. The visibility of satellites, HAPS, and gNBs is illustrated in Figure 2. From this figure, it is evident that all gNBs are in a non-line-of-sight (NLOS) position relative to the receiver. Among the HAPS, four out of six maintain a LOS with the receiver. Similarly, for satellites, four out of eight are in LOS with the receiver.

\begin{table}[t]
\caption{Associated parameters in synthetic waveform generation for satellite and HAPS.}
\begin{center}
\begin{tabularx}{0.9\linewidth}{>{\centering\arraybackslash}X|>{\centering\arraybackslash}X|>{\centering\arraybackslash}X}
\Xhline{1pt}
\textbf{Parameters} & \textbf{Satellite} & \textbf{HAPS}\\
\Xhline{1pt}
$f_s$	&\multicolumn{2}{c}{\centering 38.192 MHz}\\
\hline
$t_D$	&\multicolumn{2}{c}{\centering 20 ms}\\
\hline
$f_{IF}$	&9.548 MHz	& 15 MHz\\
\hline
$R_c$	&1.023 MHz	& 10.23 MHz\\
\Xhline{1pt}
\end{tabularx}
\label{tab2}
\end{center}
\end{table}

\section{Channel and Delay Coefficient Generation using QuaDRiGa}
In this section, we outline the enhancements implemented in our current methodology for generating channel and delay coefficients for mobile air-to-ground (A2G) links, building upon our previous work and utilizing QuaDRiGa. The simulation parameters for generating these coefficients from QuaDRiGa are detailed in Table \Romannum{1}. Refer to Table \Romannum{1}, $f_{ch}$ denotes the channel update rate, $T$ denotes the simulation period, $P_{TX}$ denotes the transmit power, $G_{TX}$ denotes the transmit antenna gain, $f_c$ denotes the carrier frequency, $M$ and $N$ denotes the number of vertical and horizontal elements, respectively, and $\phi$ denotes the electric downtilt angle of the antenna. To facilitate a fair comparison of pathloss among different ranging sources, identical transmit power and antenna gain are considered for all three ranging sources.

\begin{figure*}[t]
    \centering
    \begin{minipage}[c]{0.38\textwidth}
        \includegraphics[width=\textwidth]{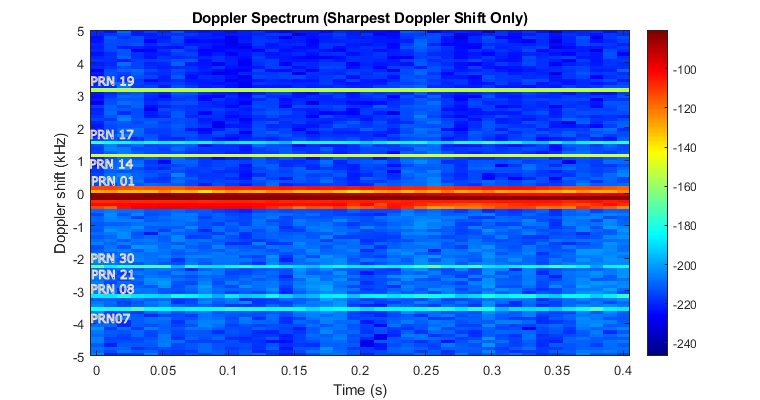}
        
        \label{fig:left}
    \end{minipage}
    \hfill 
    \hspace{2pt}
    \begin{minipage}[c]{0.6\textwidth}
        \begin{tabular}{ccccc}
            \includegraphics[height=1.9cm,width=0.18\textwidth]{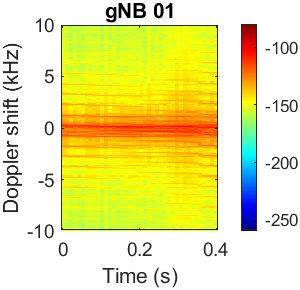} &            \includegraphics[height=1.9cm,width=0.18\textwidth]{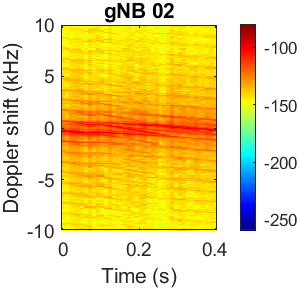} &
            \includegraphics[height=1.9cm,width=0.18\textwidth]{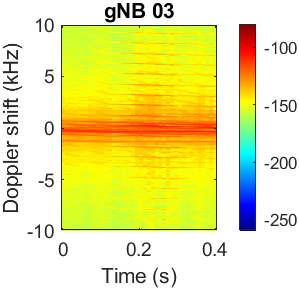} &
            \includegraphics[height=1.9cm,width=0.18\textwidth]{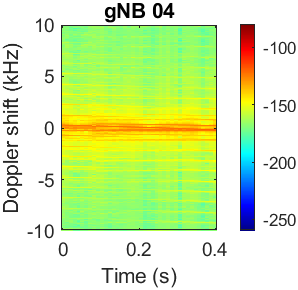} &
            \includegraphics[height=1.9cm,width=0.18\textwidth]{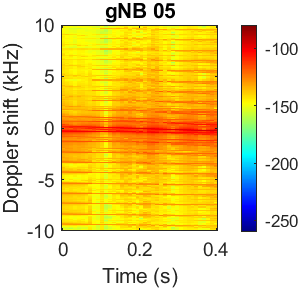} \\
            \includegraphics[height=1.9cm,width=0.18\textwidth]{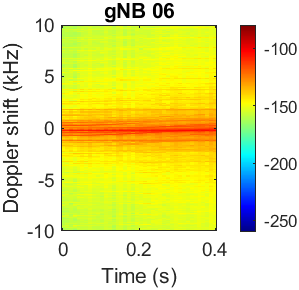} &
            \includegraphics[height=1.9cm,width=0.18\textwidth]{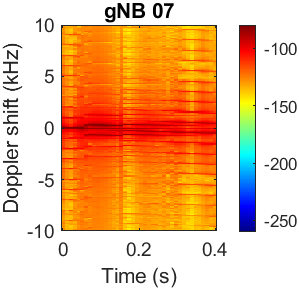} &
            \includegraphics[height=1.9cm,width=0.18\textwidth]{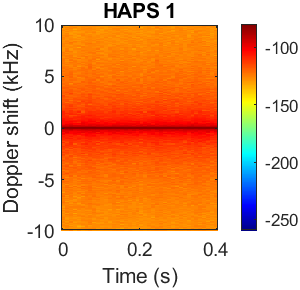} &
            \includegraphics[height=1.9cm,width=0.18\textwidth]{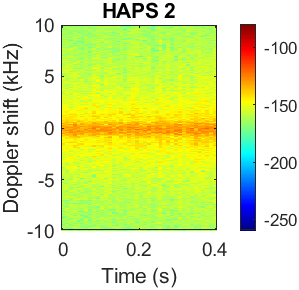} &
            \includegraphics[height=1.9cm,width=0.18\textwidth]{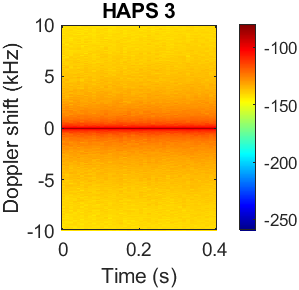} \\
            \includegraphics[height=1.9cm,width=0.18\textwidth]{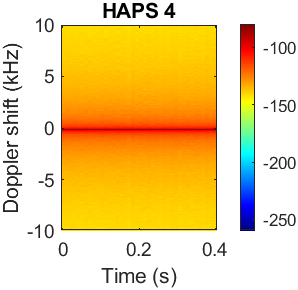} &
            \includegraphics[height=1.9cm,width=0.18\textwidth]{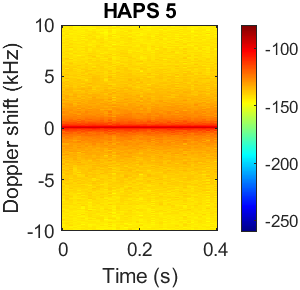} &
            \includegraphics[height=1.9cm,width=0.18\textwidth]{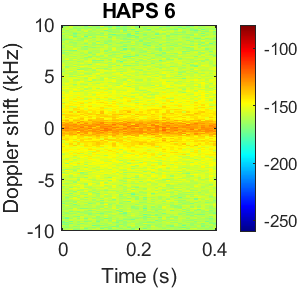} &
            \includegraphics[height=1.9cm,width=0.18\textwidth]{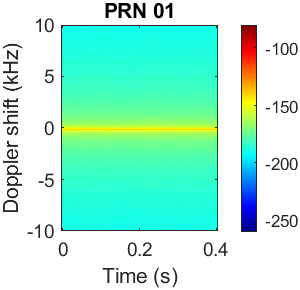} 
        \end{tabular}
        \label{fig:right}
    \end{minipage}
    \caption{Doppler spectrum for satellite, HAPS, and gNB.}
    \label{fig6}
\end{figure*}

Using the specified simulation parameters, we generate channel and delay coefficients with the QuaDRiGa channel generator. We follow the same approach as in our previous work to produce the  Doppler spectrum for satellites, HAPS, and gNBs. We consider a channel bandwidth of 2 MHz for both satellites and HAPS, and 5 MHz for gNBs. Additionally, we employ a Fast Fourier Transform (FFT) with 1024 points. The resulting Doppler spectrum is displayed in Fig.~\ref{fig6}. From this figure, it's noticeable that seven satellites exhibit Doppler shifts distinctly different from 0 Hz, yet within a range of +/-5 kHz. The Doppler shifts for HAPS and gNBs, as well as one satellite, namely pseudo-random noise (PRN) 01, are approximately 0 Hz. A more detailed view of the Doppler spectrum for these sources is provided on the right side of Fig.~\ref{fig6}. This detailed Doppler spectrum reveals that the received power from the NLOS gNBs is marginally stronger than that from the NLOS HAPS, but slightly weaker than from the LOS HAPS. The Doppler shift for HAPS is found to be similar to that for gNBs, which can be attributed to the higher carrier frequency used for gNBs and their lower elevation angles. Additionally, it is evident that the multipath effect is more pronounced for gNBs compared to HAPS and satellites.

\section{Synthetic Waveform Generation}
\subsection{Satellite and HAPS}
\begin{figure}[t]
\centerline{\resizebox{0.9\columnwidth}{!}
{\includegraphics{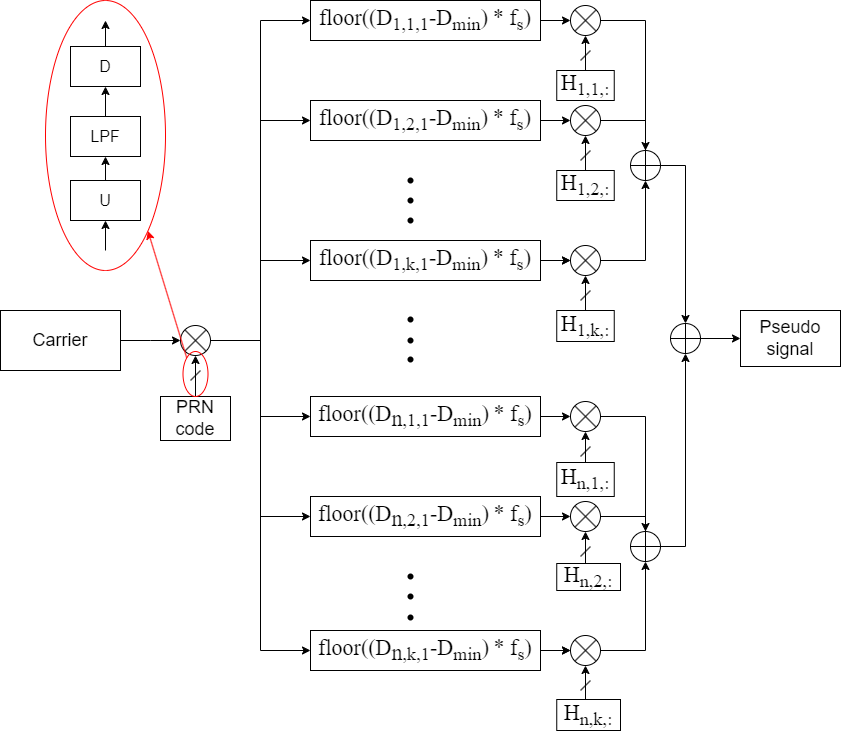}}}
\caption{Flowchart of the synthetic waveform generation for satellite and HAPS.}
\label{fig1}
\end{figure}

This section describes the process of synthetic waveform generation for satellites, HAPS, and gNBs, utilizing the channel and delay coefficients generated from QuaDRiGa channel generator. The synthetic signals for satellites and HAPS are generated in the same manner, where the code division multiple access (CDMA) is considered for signal waveform transmission and the binary phase shift keying (BPSK) is considered as the modulation scheme. The synthetic signal for gNBs is generated based on a tutorial provided by Matlab which is about the NR positioning using PRS. This guide aligns with the 3GPP specifications for 5G NR PRS, which adopt the orthogonal frequency division multiplexing (OFDM) as the signal transmission waveform. OFDM is chosen for its effectiveness and its capability to manage high data rates while reducing problems such as multipath fading and interference. For the modulation of PRS, the quadrature phase shift keying (QPSK) is utilized, providing an optimal compromise between spectral efficiency and resilience to channel impairments.

Fig.~\ref{fig1} illustrates the generation process of the synthetic signal for both satellites and HAPS. As we can see from this figure, the carrier, which is a sinusoidal wave with a predefined intermediate frequency (IF) $f_{IF}$ is firstly multiplied with the PRN code sequence at a specified chipping rate $R_c$. The red circle on the figure shows a resampling process that involves upsampling, low-pass filter (LPF), and downsampling. The main objective of the LPF is to reduce unwanted spectral repetition in signal rate change. The resultant waveform is then delayed based on each initial path delay for a satellite/HAPS. To skip the transient period, the initial delays are subtracted by the smallest initial delay $D_{min}$ among all satellites/HAPS. $D_{n,k,1}$ denotes the initial delay for the $k^{th}$ signal path from the $n^{th}$ satellite/HAPS.  Afterward, the delayed signals are multiplied by the channel coefficients. $H_{n,k,:}$ denotes the channel coefficients spanning the entire simulation period for the $k^{th}$ signal path from the $n^{th}$ satellite/HAPS. Finally, the synthetic signal is generated by summing up all the delayed signals. Table \Romannum{2} encapsulates the essential parameters utilized for generating synthetic signals for GPS satellites and HAPS. Refer to Table \Romannum{2}, $f_s$ denotes the sampling frequency and $t_D$ is the duration of a navigation bit. The selection of parameters for GPS satellites is based on the guidelines provided in~\cite{b11}.

\subsection{gNB}
\begin{figure}[t]
\centerline{\resizebox{0.9\columnwidth}{!}
{\includegraphics{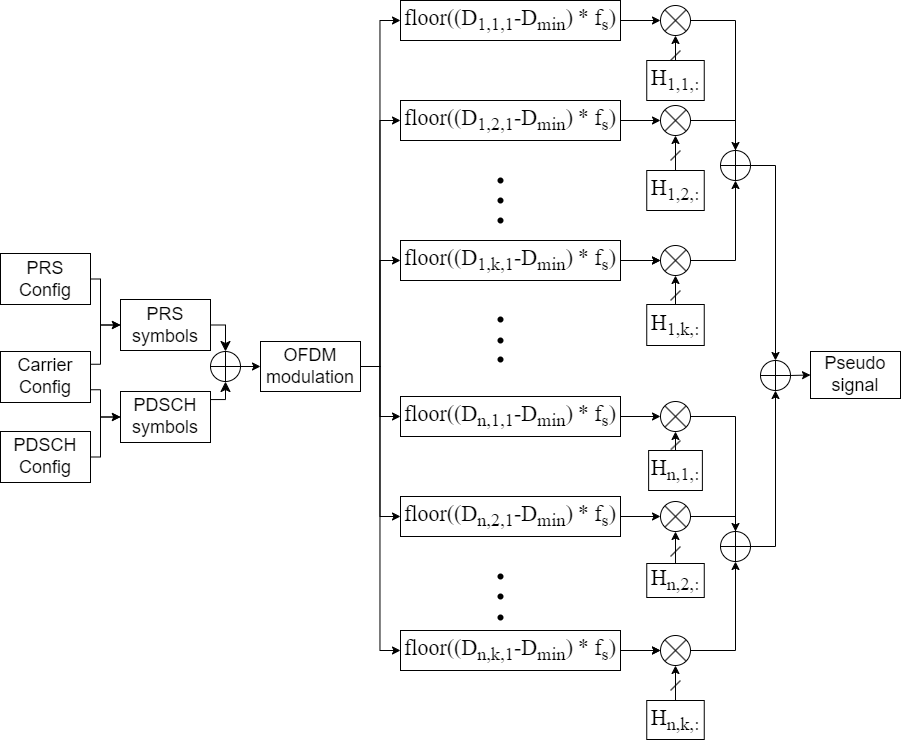}}}
\caption{Flowchart of the synthetic waveform generation for gNB.}
\label{fig2}
\end{figure}

Fig.~\ref{fig2} illustrates the generation process of the synthetic signal for gNBs. The first half of the process before the OFDM modulation is based on a tutorial provided by Matlab. As we can observe from this figure, three modules should be configured before generating PRS and physical downlink shared channel (PDSCH) symbols: PRS, carrier, and PDSCH modules. The PRS module is used to configure the PRS-associated parameters, as defined in TS 38.211 Section 7.4.1.7~\cite{b3}. These parameters include resource set period, resource offset, resource repetition, resource time gap, muting pattern, number of resource blocks $N_\textsf{RB}$, and so on. The versatility of the PRS module ensures that there is no overlapping among PRS/gNB signals, hence ensuring hearability. The carrier module is used to configure the carrier-associated parameters for a specific OFDM numerology, as defined in TS 38.211 Sections 4.2, 4.3, and 4.4~\cite{b3}. Some of the supported functions include physical layer cell identity, subcarrier spacing (SCS), and cyclic prefix. The PDSCH module is used to configure the PDSCH-associated parameters, as defined in TS 38.211 Sections 7.3.1, 7.4.1.1, and 7.4.1.2~\cite{b3}. These parameters span from the number of transmission layers, modulation scheme, demodulation reference signal (DMRS), and so forth. A detailed description of the configuration for these three modules can be found in Matlab. With PRS and PDSCH symbols along with carrier configuration, the NR OFDM modulation is then carried out to generate the OFDM modulated transmitted waveform. The waveform sample rate, $f_s$, is determined as follows:
\begin{equation}
f_s = N_\textsf{fft} \times \textsf{SCS} \label{eq1}
\end{equation}

\noindent where SCS denotes the subcarrier spacing. The OFDM modulation is carried out on the PRS and PDSCH symbols, which generates a basic waveform for a gNB. Next, the basic gNB waveform is delayed and multiplied with the channel coefficients in a similar manner as satellites and HAPS. By summing up all the delayed gNB signals, the waveform of the synthetic gNB signal is obtained. The relevant parameters used in the generation of synthetic signals for gNBs are summarized in Table \Romannum{3}. Refer to Table \Romannum{3}, $t_f$ denotes the duration of a frame in NR.

\begin{table}[t]
\caption{Associated parameters in synthetic waveform generation for gNB.}
\begin{center}
\begin{tabularx}{0.9\linewidth}{>{\centering\arraybackslash}X|>{\centering\arraybackslash}X}
\Xhline{1pt}
\textbf{Parameters} & \textbf{gNB}\\
\Xhline{1pt}
$f_s$	&15.36 MHz	\\
\hline
$f_c$	&4 GHz	\\
\hline
$t_f$	&10 ms	\\
\hline
$N_{\textsf{fft}}$	&1024	\\
\hline
$N_{\textsf{RB}}$	&52	\\
\hline
Symbols per slot	&14	\\
\hline
Slots per subframe	&1	\\
\hline
Slots per frame	&10	\\
\hline
Subcarrier spacing (SCS)	&15 kHz	\\
\Xhline{1pt}\\
\end{tabularx}
\label{tab1}
\end{center}
\end{table}

\section{Signal Acquisition and Tracking Evaluation}
As the same technique is utilized for all considered ranging sources in generating synthetic signals in the second half of the process, where the generated channel and delay coefficients are utilized, the signal acquisition and tracking results are presented only for synthetic satellite signals. The goal is to validate the generated synthetic signals in the sense that the code delay and Doppler shift can be accurately estimated. We consider the commonly used parallel code phase searching algorithm for signal acquisition, and the conventional tracking algorithm for signal tracking~\cite{b11}.

\subsection{Acquisition}
\begin{figure*}[t]
\centering
\centering
\begin{tabularx}{0.68\textwidth}{>{\centering\arraybackslash}X >{\centering\arraybackslash}X}
\includegraphics[height=4.24cm,width=\linewidth]{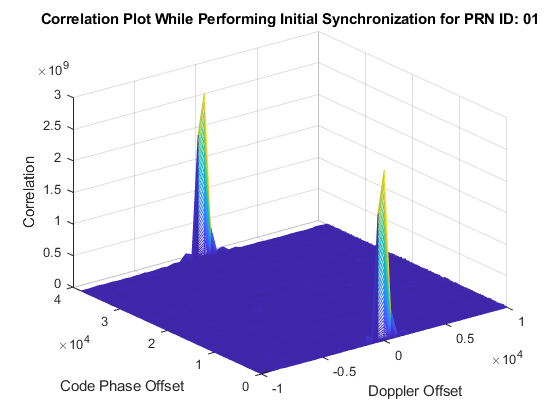}&
\includegraphics[height=4.24cm,width=\linewidth]{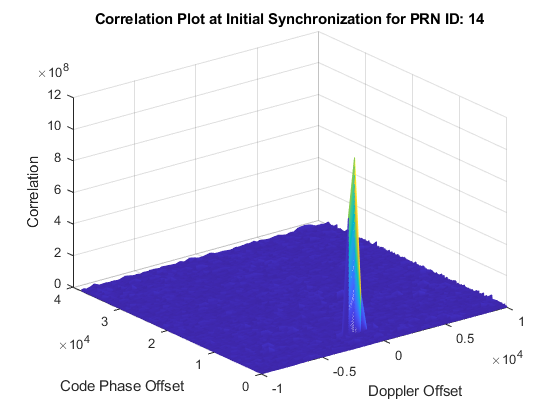}\\
\includegraphics[height=4.24cm,width=\linewidth]{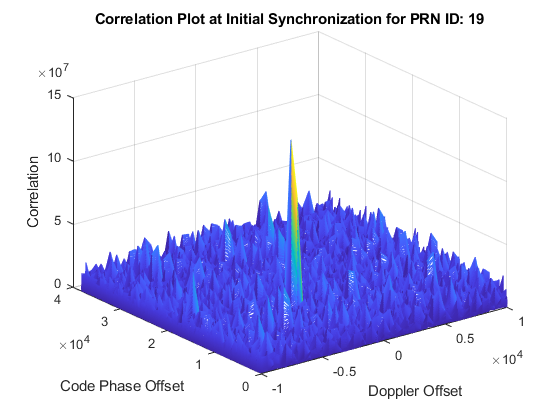}&
\includegraphics[height=4.24cm,width=\linewidth]{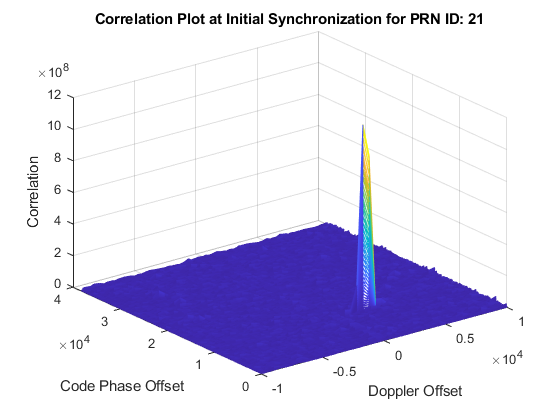}\\
\end{tabularx}
\caption{Correlation plot for PRN 01, PRN 14, PRN 19, and PRN 21.}
\label{fig8}
\end{figure*}

The parallel code phase search algorithm is employed to determine the estimated code delay $\hat{\tau}$ and carrier frequency $\hat{f_c}$ of satellite signals. To exclude low-quality signals, the signal-to-noise ratio (\textsf{SNR}) is utilized as a filtering criterion. The SNR is calculated as follows:
\begin{equation}
\textsf{SNR} = 10\log_{10}(\frac{r_{\mathrm{max},\hat{f_c}}^2}{\frac{1}{L}\sum_i^{N_cN_{s}}r_{i,\hat{f_c}}^2}), \forall i \notin (\hat{\tau}-N_s,\hat{\tau}+N_s) \label{eq1}
\end{equation}

\noindent where $r_{i,\hat{f_c}}$ is the correlation value along the estimated carrier frequency, $r_{\mathrm{max},\hat{f_c}}$ denotes the correlation peak, $L$ denotes the length of the noise samples, $N_c$ denotes the number of chips for a complete cycle of the spreading code, which is 1023 for GPS L1 C/A code, and $N_s$ denotes the number of samples for a chip at the sample rate we consider. The SNR threshold is 25 dB. The typical frequency search step in GPS L1 signal acquisition is 500 Hz, with the estimated code delay, the Doppler frequency shift is estimated by the fine frequency calculation where a longer signal is analyzed. In general, the longer the signal is analyzed, the more accurate the Doppler shift can be estimated, while the longer computation time it takes. In this work, 10 ms signal is used for fine frequency calculation. The correlation plots for the acquired satellites PRN 01, PRN14, PRN 19, and PRN 21 are presented in Fig.~\ref{fig8}. We notice that all the acquired satellites have LOS with the receiver. The NLOS satellites are not acquired successfully due to their low SNR and channel fading.

\subsection{Tracking}
\begin{figure*}[t]
\centering
\begin{tabularx}{\textwidth}{>{\centering\arraybackslash}X >{\centering\arraybackslash}X >{\centering\arraybackslash}X >{\centering\arraybackslash}X}
\includegraphics[height=2.1cm,width=\linewidth]{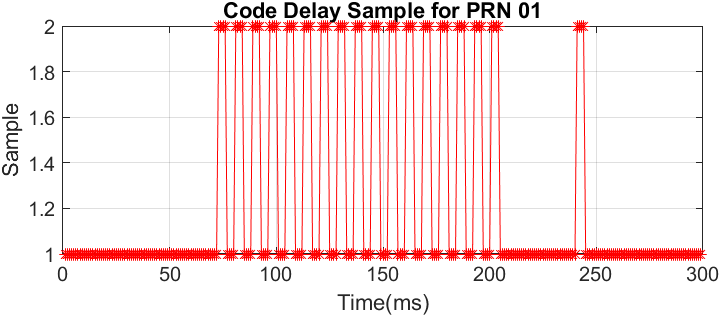}&
\includegraphics[height=2.1cm,width=\linewidth]{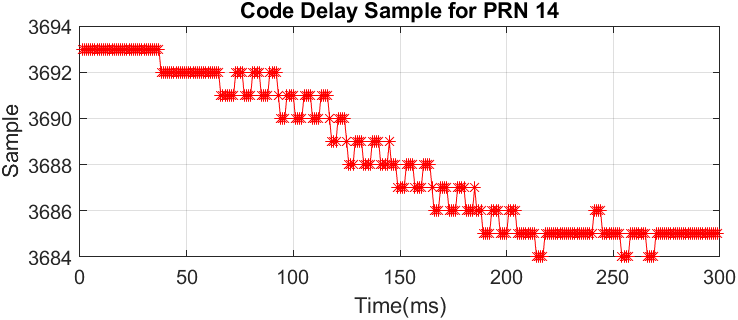}&
\includegraphics[height=2.1cm,width=\linewidth]{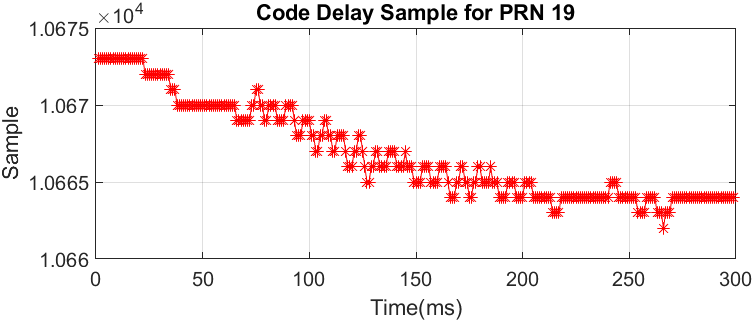}&
\includegraphics[height=2.1cm,width=\linewidth]{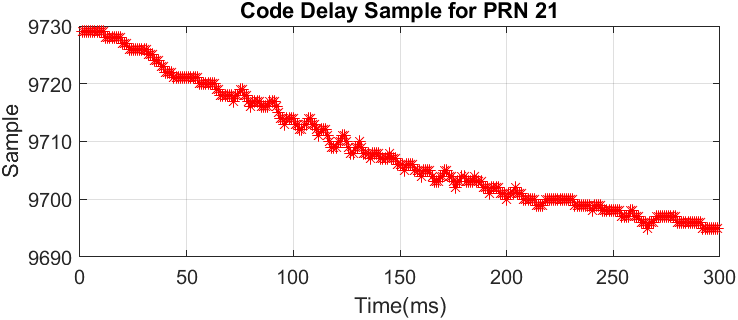}\\
\includegraphics[height=2.1cm,width=\linewidth]{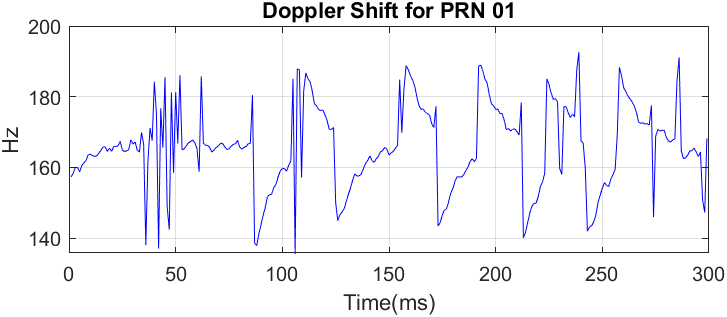}&
\includegraphics[height=2.1cm,width=\linewidth]{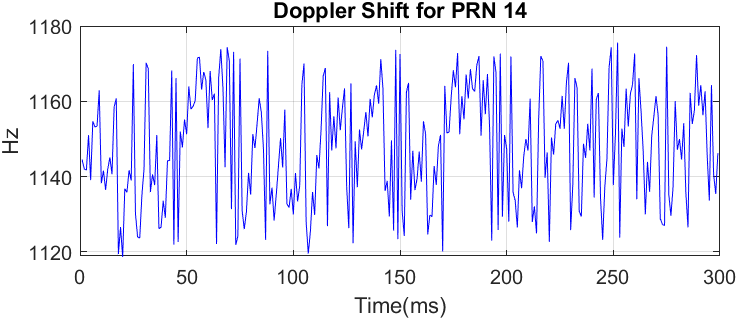}&
\includegraphics[height=2.1cm,width=\linewidth]{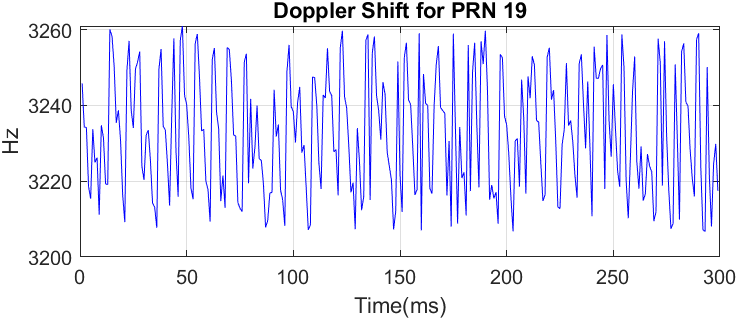}&
\includegraphics[height=2.1cm,width=\linewidth]{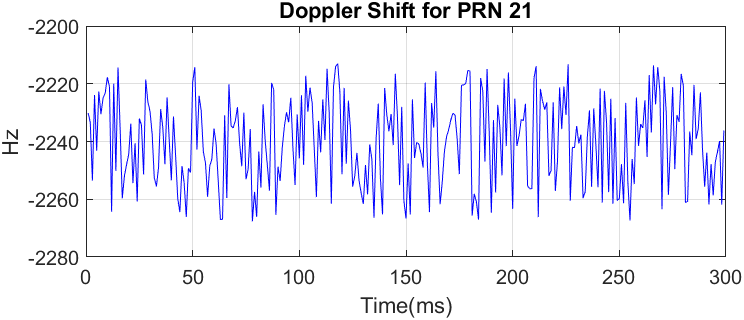}\\
\end{tabularx}
\caption{Estimated code delay and Doppler shift for PRN 01, PRN 14, PRN 19, and PRN 21.}
\label{fig3}
\end{figure*}

The conventional tracking loop used for GPS satellite signals is considered when tracking the acquired satellites. The normalized early-minus-late power is used to calculate the loop discriminator. Fig.~\ref{fig3} presents the estimated code delay and Doppler shift for the acquired satellites. From this figure, we can observe that the estimated Doppler shifts for the acquired satellites are roughly consistent with that shown in the Doppler spectrum in Fig.~\ref{fig6}. We verify that the estimated initial code delay samples roughly match the expected values. For example, the delay of PRN 01 is the smallest among all satellites, hence a zero code delay is expected for this satellite. It can be noted that the fluctuation in the Doppler shift is quite large, this is because we are using the common choice of the parameters in the tracking loop, such as a 2 Hz noise bandwidth for the delay-lock loop (DLL), a 10 Hz noise bandwidth for the phase-lock loop (PLL), and so on. Since the choice of tracking parameters depends on the specific scenario, the tracking performance is expected to improve if we consider using adaptive loop-bandwidth tracking techniques such as fast adaptive bandwidth (FAB), fuzzy logic (FL), and loop-bandwidth control algorithm (LBCA) proposed in~\cite{b12}.

\section{Conclusion}
In this study,  we demonstrate that generating a realistic signal waveform is feasible by utilizing channel and delay coefficients in conjunction with our proposed methodology. These coefficients are generated from a well recognized channel generator that aligns with 3GPP standards, namely QuaDRiGa. Since the method for incorporating these channel and delay coefficients into the waveform generation for satellite, HAPS, and gNB PRS is consistent, the precise estimation of code delay and Doppler shift for satellite suggests that the synthetic waveform generation for HAPS and gNB PRS is equally effective. This research marks a significant step forward in the field of waveform generation and heterogeneous navigation systems, particularly for the envisioned objects. Additionally, we believe that the implementation of an adaptive loop-bandwidth tracking technique could significantly improve tracking performance.

\vspace{12pt}


\begin{thebibliography}{00}
\bibitem{b4} G. K. Kurt, M. G. Khoshkholgh, S. Alfattani, A. Ibrahim, T. S. J. Darwish, M. S. Alam, H. Yanikomeroglu, and A. Yongacoglu, ``A vision and framework for the high altitude platform station (HAPS) networks of the future,'' \textit{IEEE Communications Surveys \& Tutorials}, vol. 23, no. 2, pp. 729-779, Secondquarter 2021.

\bibitem{b16} S. Dwivedi, R. Shreevastav, F. Munier, J. Nygren, I. Siomina, Y. Lyazidi, D. Shrestha, G. Lindmark, P. Ernström, E. Stare, S. M. Razavi, S. Muruganathan, G. Masini, Å. Busin, and F. Gunnarsson, ``Positioning in 5G networks,'' \textit{IEEE Communications Magazine}, vol. 59, no. 11, pp. 38-44, Nov. 2021.

\bibitem{b5} H. -H. Cho, C. -F. Lai, T. K. Shih, and H. -C. Chao, ``Integration of SDR and SDN for 5G,'' \textit{IEEE Access}, vol. 2, pp. 1196-1204, 2014.

\bibitem{b6} D. Sinha, A. K. Verma, and S. Kumar, ``Software defined radio: Operation, challenges and possible solutions,'' in \textit{Proceedings of 10th International Conference on Intelligent Systems and Control (ISCO)}, Coimbatore, India, 2016, pp. 1-5. 

\bibitem{b7} M. Kurras, S. Dai, S. Jaeckel, and L. Thiele, ``Evaluation of the spatial consistency feature in the 3GPP geometry-based stochastic channel model,'' in \textit{Proceedings of IEEE Wireless Communications and Networking Conference (WCNC)}, Marrakesh, Morocco, 2019, pp. 1-6.

\bibitem{b8} J. Medbo, K. Börner, K. Haneda, V. Hovinen, T. Imai, J. Järvelainen, T. Jämsä, A. Karttunen, K. Kusume, J. Kyröläinen, P. Kyösti, J. Meinilä, V. Nurmela, L. Raschkowski, A. Roivainen, and J. Ylitalo, ``Channel modelling for the fifth generation mobile communications,'' in \textit{Proceedings of 8th European Conference on Antennas and Propagation (EuCAP 2014)}, The Hague, Netherlands, 2014, pp. 219-223.

\bibitem{b9} C. -X. Wang, J. Bian, J. Sun, W. Zhang, and M. Zhang, ``A survey of 5G channel measurements and models,'' \textit{IEEE Communications Surveys \& Tutorials}, vol. 20, no. 4, pp. 3142-3168, Fourthquarter 2018.

\bibitem{b17} S. Jaeckel, L. Raschkowski, K. Börner, and L. Thiele, ``QuaDRiGa: A 3-D multi-cell channel model with time evolution for enabling virtual field trials,'' \textit{IEEE Transactions on Antennas and Propagation}, vol. 62, no. 6, pp. 3242-3256, Jun. 2014.

\bibitem{b18} I. Burtakov, A. Kureev, A. Tyarin, and E. Khorov, ``QRIS: A QuaDRiGa-based simulation platform for reconfigurable intelligent surfaces,'' \textit{IEEE Access}, vol. 11, pp. 90670-90682, 2023.

\bibitem{b19} S. Jaeckel, L. Raschkowski, and L. Thieley, ``A 5G-NR satellite extension for the QuaDRiGa channel model,'' in \textit{Proceedings of Joint European Conference on Networks and Communications \& 6G Summit (EuCNC/6G Summit)}, Grenoble, France, 2022, pp. 142-147.

\bibitem{b20} S. Dai, N. Abdellatif, M. Kurras, S. Jaeckel, and L. Thiele, ``Spatial consistency validation on massive SIMO covariance matrices in the geometry-based stochastic channel model QuaDRiGa,'' in \textit{Proceedings of WSA 2020; 24th International ITG Workshop on Smart Antennas}, Hamburg, Germany, 2020, pp. 1-6.

\bibitem{b2} H. Zheng, M. Atia, and H. Yanikomeroglu, ``Realistic channel and delay coefficient generation
for dual mobile space-ground links – A tutorial,'' under review in \textit{IEEE Open Journal of Vehicular Technology}.

\bibitem{b10} H. Zheng, M. Atia, and H. Yanikomeroglu, ``A positioning system in an urban vertical heterogeneous network (VHetNet),'' \textit{IEEE Journal of Radio Frequency Identification}, vol. 7, pp. 352-363, Apr. 2023.

\bibitem{b13} F. Hsieh and M. Rybakowski, ``Propagation model for high altitude platform systems based on ray tracing simulation,'' in \textit{Proceedings of 13th European Conference on Antennas and Propagation (EuCAP)}, Krakow, Poland, 2019, pp. 1-5.

\bibitem{b14} S. Alfattani, W. Jaafar, Y. Hmamouche, H. Yanikomeroglu, and A. Yongacoglu, ``Link budget analysis for reconfigurable smart surfaces in aerial platforms,'' \textit{IEEE Open Journal of the Communications Society}, vol. 2, pp. 1980-1995, 2021.

\bibitem{b15} 3GPP TR 38.901, ``5G; Study on channel model for frequencies from 0.5 to 100 GHz (3GPP TR 38.901 version 16.1.0 Release 16),'' 3rd Generation Partnership Project, 2020. 

\bibitem{b1} 3GPP TS 38.215, ``NR; Physical layer measurements (Release 16),'' 3rd Generation Partnership Project, Technical Specification Group Radio Access Network, Mar. 2023.

\bibitem{b11} K. Borre, D. M. Akos, N. Bertelsen, P. Rinder, and S. H. Jensen, \textit{A software-defined GPS and Galileo receiver: A single-frequency approach}, Applied and Numerical Harmonic Analysis (ANHA), Boston, US, Birkhäuser, 2006.

\bibitem{b3} 3GPP TS 38.211, ``NR; Physical channels and modulation (Release 16),'' 3rd Generation Partnership Project, Technical Specification Group Radio Access Network, Jul. 2020.

\bibitem{b12} I. Cortes, J. R. van der Merwe, J. Nurmi, A. Rügamer, and W. Felber, ``Evaluation of adaptive loop-bandwidth tracking techniques in GNSS receivers,'' \textit{Sensors}, vol. 21, no. 2, pp. 502, Jan. 2021.



\end{thebibliography}
\end{document}